\documentclass[final,3p,times,twocolumn]{elsarticle}

\usepackage{amssymb}
\usepackage{amsmath}

\usepackage{algorithmicx}
\usepackage{algpseudocode}
\usepackage{algorithm}
\algnewcommand\algorithmicinput{\textbf{INPUT:}}
\algnewcommand\INPUT{\item[\algorithmicinput]}

\algnewcommand\algorithmicoutput{\textbf{OUTPUT:}}
\algnewcommand\OUTPUT{\item[\algorithmicoutput]}

\usepackage{multirow}
\usepackage[binary-units=true]{siunitx}

\usepackage{url}

\journal{Future Generation Computer Systems}

\begin{document}

\begin{frontmatter}



\title{An efficient cloud scheduler design supporting preemptible
    instances\tnoteref{doi}}

\tnotetext[doi]{This is the author's accepted version of the following article:
    {\'A}lvaro L{\'o}pez Garc{\'i}a, Enol Fernández del Castillo, Isabel Campos Plasencia,
    ``An efficient cloud scheduler design supporting preemptible instances'',
    accepted in Future Generation Computer Systems, 2019, which is published in
    its final form at \url{https://doi.org/10.1016/j.future.2018.12.057}. This
    preprint article may be used for non-commercial purposes under a CC
    BY-NC-SA 4.0 license.
}


\author{{\'A}lvaro L{\'o}pez Garc{\'i}a\corref{cor1}}

\cortext[cor1]{Corresponding author}
\ead{aloga@ifca.unican.es}

\author{Enol Fern{\'a}ndez del Castillo}
\ead{enolfc@ifca.unican.es}

\author{Isabel Campos Plasencia}
\ead{iscampos@ifca.unican.es}

\address{Institute of Physics of Cantabria, Spanish National Research Council --- IFCA (CSIC---UC).\\
Avda. los Castros s/n. 39005 Santander, Spain}

\begin{abstract}
    Maximizing resource utilization by performing an efficient resource
    provisioning is a key factor for any cloud provider: commercial actors can
    maximize their revenues, whereas scientific and non-commercial providers
    can maximize their infrastructure utilization. Traditionally, batch systems
    have allowed data centers to fill their resources as much as possible by
    using backfilling and similar techniques.  However, in an IaaS cloud, where
    virtual machines are supposed to live indefinitely, or at least as long as
    the user is able to pay for them, these policies are not easily
    implementable. In this work we present a new scheduling algorithm for IaaS
    providers that is able to support preemptible instances, that can be stopped
    by higher priority requests without introducing large modifications in the
    current cloud schedulers. This scheduler enables the implementation of new
    cloud usage and payment models that allow more efficient usage of the
    resources and potential new revenue sources for commercial providers. We
    also study the correctness and the performace overhead of the proposed
    scheduler agains existing solutions.
\end{abstract}

\begin{keyword}
Cloud computing \sep scheduling \sep preemptible instances \sep Spot Instances \sep resource allocation



\end{keyword}

\end{frontmatter}

\section{Introduction}
\label{sec:introduction}

Infrastructure as a Service (IaaS) Clouds make possible to provide computing
capacity as a utility to the users following a pay-per-use model. This fact
allows the deployment of complex execution environments without an upfront
infrastructure commitment, fostering the adoption of the cloud by users that
could not afford to operate an on-premises infrastructure. In this regard,
Clouds are not only present in the industrial ICT ecosystem, and they are being
more and more adopted by other stakeholders such as public administrations or
research institutions.

Indeed, clouds are nowadays common in the scientific computing field
\cite{Hoffa2008,Iosup2011,Vockler2011,LopezGarcia2017c}, due to the fact that
they are able to deliver resources that can be configured with the complete
software needed for an application \cite{Juve2008}. Moreover, they
also allow the execution of non-transient tasks, making possible to execute
virtual laboratories, databases, etc. that could be tightly coupled with the
execution environments. This flexibility poses a great advantage against
traditional computational models ---such as batch systems or even Grid
computing--- where a fixed operating system is normally imposed and any
complimentary tools (such as databases) need to be self-managed outside the
infrastructure. This fact is pushing scientific datacenters outside their
traditional boundaries, evolving into a mixture of services that deliver more
added value to their users, with the Cloud as a prominent actor.

Maximizing resource utilization by performing an efficient resource
provisioning is a fundamental aspect for any resource provider, specially for
scientific providers. Users accessing these computing resources do not usually
pay ---or at least they are not charged directly--- for their consumption, and
normally resources are paid via other indirect methods (like access grants),
with users tending to assume that resources are \emph{for free}. Scientific
computing facilities tend to work on a fully saturated manner, aiming at the
maximum possible resource utilization level.

In this context it is common that compute servers spawned in a cloud
infrastructure are not terminated at the end of their lifetime, resulting in
idle resources, a state that is are not desirable as long as there is
processing that needs to be done \cite{LopezGarcia2017c}. In a commercial this
is not a problem, since users are being charged for their allocated resources,
regardless if they are being used or not. Therefore users tend to take care of
their virtual machines, terminating them whenever they are not needed anymore.
Moreover, in the cases where users leave their resources running forever, the
provider is still obtaining revenues for those resources.

Cloud operators try to solve this problem by setting resource quotas that
limits the amount of resources that a user or group is able to consume by doing
a static partitioning of the resources \cite{LopezGarcia2017}. However, this
kind of resource allocation automatically leads to an underutilization of the
infrastructure since the partitioning needs to be conservative enough so that
other users could utilize the infrastructure. Quotas impose hard limits that
leading to dedicated resources for a group, even if the group is not using the
resources.

Besides, cloud providers also need to provide their users with on-demand access
to the resources, one of the most compelling cloud characteristics
\cite{Ramakrishnan2011}. In order to provide such access, an overprovisioning
of resources is expected \cite{Marshall2011} in order to
fulfil user request, leading to an infrastructure where utilization is not
maximized, as there should be always enough resources available for a potential
request.

Taking into account that some processing workloads executed on the cloud do not
really require on-demand access (but rather they are executed for long periods
of time), a compromise between these two aspects (i.e.  maximizing utilization
and providing enough on-demand access to the users) can be provided by using
idle resources to execute these tasks that do not require truly on-demand
access \cite{Marshall2011}. This approach indeed is
common in scientific computing, where batch systems maximize the resource
utilization through \emph{backfilling} techniques, where opportunistic access
is provided to these kind of tasks.

Unlike in batch processing environments, virtual machines (VMs) spawned in a
Cloud do not have fixed duration in time and are supposed to live forever ---or
until the user decides to stop them. Commercial cloud providers provide
specific VM types (like the Amazon EC2 Spot Instances\footnote{\url{http://aws.amazon.com/ec2/purchasing-options/spot-instances/}}
or the Google Compute Engine Preemptible Virtual Machines\footnote{\url{https://cloud.google.com/preemptible-vms/}}) that can be
provisioned at a fraction of a normal VM price, with the caveat that they can
terminated whenever the provider decides to do so. This kind of VMs can be used
to backfill idle resources, thus allowing to maximize the utilization and
providing on-demand access, since normal VMs will obtain resources by
evacuating Spot or Preemptible instances.

In this paper we propose an efficient scheduling algorithm that combines the
scheduling of preemptible and non preemptible instances in a modular way. The
proposed solution is flexible enough in order to allow different allocation,
selection and termination policies, thus allowing resource providers to easily
implement and enforce the strategy that is more suitable for their needs.
In our work we extend the OpenStack Cloud middleware with a prototype
implementation of the proposed scheduler, as a way to demonstrate and evaluate
the feasibility of our solution. We moreover perform an evaluation of the
performance of this solution, in comparison with the existing OpenStack
scheduler.

The remainder of the paper is structured as follows.
In Section~\ref{sec:related} we present the related work in this field.
In Section~\ref{sec:design} we propose a design for an efficient scheduling
mechanism for preemptible instances.
In Section~\ref{sec:evaluation} we present an implementation of our proposed
algorithm, as well as an evaluation of its feasibility and performance with
regards with a normal scheduler.
Finally, in Section~\ref{sec:conclusions} we present this work's conclusions.

\section{Related work}
\label{sec:related}

The resource provisioning from cloud computing infrastructures using Spot
Instances or similar mechanisms has been addressed profusely in the scientific
literature in the last years \cite{Jennings2015}. However, the vast majority of
this work has been done from the users' perspective when using and consuming
Spot Instances \cite{DeAssuncao2016} and few works tackle the problem from the
resource provider standpoint.

Due to the unpredictable nature of the Spot Instances, there are several
research papers that try to improve the task completion time ---making the task
resilient against termination--- and reduce the costs for the user. Andrzejak
et al. \cite{Andrzejak2010} propose a probabilistic model to obtain the bid
prices so that the costs and performance and reliability can be improved. In
\cite{Yi2010,Yi2012,Khatua2013,Jung2011} the task checkpointing is addressed so
as to minimize costs and improve the whole completion time.

Related with the previous works, Voorsluys et al. have studied the usage of
Spot Instances to deploy reliable virtual clusters
\cite{Voorsluys2011,Voorsluys2012}, managing the allocated instances on
behalf of the users. They focus on the execution of compute intensive tasks on
top of a pool of Spot Instances, in order to find the most effective way to
minimize both the execution time of a given workload and the price of the
allocated resources. Similarly, in \cite{jung2014workflow} the autors develop a
workflow scheduling scheme that reduces the completion time using Spot
Instances.

Jain et al. have performed studies in the same line, but focused on using a
batch system that leverages the Spot Instances \cite{Jain2014}, learning from
its previous experience ---in terms of spot prices and workload
characteristics--- in order to dynamically adapt the resource allocation
policies of the batch system.

Regarding Big Data analysis, several authors have studied how the usage of Spot
Instances could be used to execute MapReduce workloads reducing the monetary
costs, such as in \cite{Chohan2012,Liu2011}. The usage of Spot Instances for
opportunistic computing is another usage that has awaken a lot of interest,
especially regarding the design of an optimal bidding algorithm that would
reduce the costs for the users \cite{6195567,6481231}. There are already
existing applications such as the vCluster framework \cite{Noh2013} that can
consume resources from heterogeneous cloud infrastructures in a fashion that
could take advantage of the lower price that the Spot Instances should provide.

In spite of the above works, to the best of our knowledge, there is a lack of
research in the feasibility, problematic, challenges and implementation from
the perspective of the IaaS provider. In spite of the user's interest in
exploiting preemptible instances and the large commercial actors providing this
alternative payment and access model, it is hard to find open source products
or implementations of preemptible instances.

Amazon provides the EC2 Spot
Instances\footnote{\url{http://aws.amazon.com/ec2/purchasing-options/spot-instances/}},
where users are able to select how much they are willing to pay for their
resources by \emph{bidding} on their price in market where the price fluctuates
accordingly to the demand. Those requests will be executed taking into account
the following points:

\begin{itemize}
    \item The EC2 Spot Instances will run as long as the published Spot price
        is lower than their bid.
    \item The EC2 Spot Instance will be terminated when the Spot price is higher
        than the bid (out-of-bid).
    \item If the user terminates the Spot Instance, the complete usage will be
        accounted, but if it gets terminated by the system, the last partial
        hour won't be accounted.
\end{itemize}

When an out-of-bid situation happens, the running instances will be terminated
without further advise. This rough explanation of the Amazon's Spot Instances
can be considered similar to the traditional job preemption based on
priorities, with the difference that the priorities are being driven by an
economic model instead by the usual fair-sharing or credit mechanism used in
batch systems.

Google Cloud Engine
(GCE)\footnote{\url{https://cloud.google.com/products/compute-engine}} has
released a new product branded as \emph{Preemptible Virtual Machines}
\footnote{\url{https://cloud.google.com/preemptible-vms/}}. These new Virtual
Machine (VM) types are short-lived compute instances suited for batch
processing and fault-tolerant jobs, that can last for up to \SI{24}{\hour} and
that can be terminated if there is a need for more space for higher priority
tasks within the GCE.

Marshall et al. \cite{Marshall2011} delivered an
implementation of preemptible instances for the Nimbus toolkit in order to
utilize those instances for backfilling of idle resources, focusing on  HTC
fault-tolerant tasks. However, they did not focus on offering this
functionality to the end-users, but rather to the operators of the
infrastructure, as a way to maximize their resource utilization. In this work,
it was the responsibility of the provider to configure the backfill tasks that
were to be executed on the idle resources.

Nadjaran Toosi et al. have developed a Spot Instances as a Service (SIPaaS)
framework, a set of web services that makes possible to run a Spot market on
top of an OpenStack cloud \cite{NadjaranToosi2015}. However, even if this
framework aims to deliver preemptible instances on OpenStack cloud, it is
designed to utilize normal resources to provide this functionality. SIPaaS
utilizes normal resources to create the Spot market that is provided to the
users by means of a thin layer on top of a given OpenStack, providing a
different API to interact with the resources. From the CMF point of view, all
resources are of the same type, being SIPaaS the responsible of handling them,
in different ways. In contrast, our work leverages two different kind of
instances at the CMF level, performing different scheduling strategies
depending on which kind of resource it is being requested. SIPaaS also delivers
a price market similar to the Amazon EC2 Spot Instances market, therefore they
also provide the Ex-CORE auction algorithm \cite{toosi2016auction} in order to
govern the price fluctuations.

Carvalho et al. have proposed \cite{Carvalho2017} a capacity planning method
combined with an admission service for IaaS cloud providers offering different
service classes. This method allows providers to tackle the challenge of
estimating the minimum capacity required to deliver an agreed Service Level
Objective (SLO) across all the defined service classes. In the aforementioned
paper Carvalho et al. lean on their previous work
\cite{Carvalho2014,Carvalho2016}, where they proposed a way to reclaim unused
cloud resources to offer a new \emph{economy} class. This class, in contrast
with the preemptible instances described here, still offer a SLO to the users,
being the work on Carvalho et al. focused on the reduction of the changes that
the SLO is violated due to an instance reclamation because of a capacity
shortage.

\subsection{Scheduling in the existing Cloud Management Frameworks}
\label{sec:sched:algo}

Generally speaking, existing Cloud Management Frameworks (CMFs) do not
implement full-fledged queuing mechanism as other computing models do (like the
Grid or traditional batch systems). Clouds are normally  more focused on the
rapid scaling of the resources rather than in batch processing, where systems
are governed by queuing systems \cite{Foster2008}. The default scheduling
strategies in the current CMFs are mostly based on the immediate allocation or
resources following a fist-come, first-served basis. The cloud schedulers
provision them when requested, or they are not provisioned at all (except in
some CMFs that implement a FIFO queuing mechanism) \cite{buyya2010cloud}.

However, some users require for a queuing system ---or some more advanced
features like advance reservations--- for running virtual machines.  In those
cases, there are some external services such as Haizea \cite{Sotomayor2009} for
OpenNebula or Blazar \footnote{\url{https://launchpad.net/blazar}} for OpenStack. Those systems lay between
the CMF and the users, intercepting their requests and interacting with
the cloud system on their behalf, implementing the required functionality.

Besides simplistic scheduling policies like first-fit or random chance node
selection \cite{buyya2010cloud}, current CMF implement a scheduling algorithm
that is based on a rank selection of hosts, as we will explain in what follows:

\begin{description}
    \item[OpenNebula] \footnote{\url{http://opennebula.org/}} uses by default a
        \texttt{match making} scheduler, implementing the Rank Scheduling
        Policy \cite{Sotomayor2009}. This policy first performs a filtering of
        the existing hosts, excluding those that do not meet the request
        requirements. Afterwards, the scheduler evaluates some operator defined
        rank expressions against the recorded information from each of the
        hosts so as to obtain an ordered list of nodes. Finally, the resources
        with a higher rank are selected to fulfil the request. OpenNebula
        implements a queue to hold the requests that cannot be satisfied
        immediately, but this queuing mechanism follows a FIFO logic, without
        further priority adjustment.

    \item[OpenStack] \footnote{\url{http://www.openstack.org}} implements a
        Filter Scheduler \cite{Litvinski2013}, based on two separated phases.
        The first phase consists on the filtering of hosts that will exclude
        the hosts that cannot satisfy the request. This filtering follows a
        modular design, so that it is possible to filter out nodes based on the
        user request (RAM, number of vCPU), direct user input (such as instance
        affinity or anti-affinity) or operator configured filtering. The second
        phase consists on the weighing of hosts, following the same modular
        approach.  Once the nodes are filtered and weighed, the best candidate
        is selected from that ordered set.

    \item[CloudStack] \footnote{\url{https://cloudstack.apache.org}} utilizes
        the term \emph{allocator} to determine which host will be selected to
        place the new VM requested.  The nodes that are used by the allocators
        are the ones that are able to satisfy the request.

    \item[Eucalyptus] \footnote{\url{https://www.eucalyptus.com/}} implements a
        greedy or round robin algorithm. The former strategy uses the first
        node that is identified as suitable for running the VM. This algorithm
        exhausts a node before moving on to the next node available. On the
        other hand, the later schedules each request in a cyclic manner,
        distributing evenly the load in the long term.
\end{description}

\begin{algorithm}[htb!]
    \caption{Scheduling Algorithm.}
    \label{alg:sched}
    \begin{algorithmic}[1]
        \Function{Schedule Request}{$req, H$}
            \INPUT{$req$: user request}
            \INPUT{$H$: all host states}
            \State {$hosts \gets [~]$} \Comment{empty list}
            \ForAll {$h_i \in H$}
                \If {\Call{Filter}{$h_i, req$}}
                    \State $\Omega_i \gets 0$
                    \ForAll {$r, m$ in ranks} \Comment{$r$ is a
                    rank function, $m$ the rank multiplier}
                        \State $\Omega_i \gets \Omega_i + m_j * r_j(h_i, req) $
                    \EndFor
                    \State $hosts \gets hosts + (h_i, \Omega_i)$ \Comment{append to the list}
                \EndIf
            \EndFor
            \State \Return $hosts$
        \EndFunction
    \end{algorithmic}
\end{algorithm}

All the presented scheduling algorithms share the view that the nodes are
firstly filtered out ---so that only those that can run the request are
considered--- and then ordered or ranked according to some defined rules.
Generally speaking, the scheduling algorithm can be expressed as the
pseudo-code in the Algorithm~\ref{alg:sched}.

\section{Preemptible Instances Design}
\label{sec:design}

The initial assumption for a \emph{preemptible aware} scheduler is that the
scheduler should be able to take into account two different instance types
---preemptible and normal--- according to the following basic rules:

\begin{itemize}
    \item If it is a normal instance and there are no free resources for it, it
        must check if the termination of any running preemptible instance will
        leave enough space for the new instance.
        \begin{itemize}
            \item If this is true, those instances should be terminated
                ---according to some well defined rules--- and the new VM
                should be scheduled into that freed node.
            \item If this is not possible, then the request should continue
                with the failure process defined in the scheduling algorithm
                ---it can be an error, or it can be retried after some elapsed
                time.
        \end{itemize}
    \item If it is a preemptible instance, it should try to schedule it without
        other considerations.
\end{itemize}

It should be noted that the preemptible instance selection and termination does
not only depend on pure theoretical aspects, as this selection will have an
influence on the resource provider revenues and the service level agreements
signed with their users. Taking this into account, it is obvious that
modularity and flexibility for the preemptible instance selection and is a key
aspect here. For instance, an instance selection and termination algorithm that
is only based on the minimization of instances terminated in order to free
enough resources may not work for a provider that wish to terminate the
instances that generate less revenues, event if it is needed to terminate a
larger amount of instances.

Therefore, the aim of our work is not only to design an scheduling algorithm,
but also to design it as a modular system so that it would be possible to
create any more complex model on top of it once the initial preemptible
mechanism is in place.

The most evident design approach is a retry mechanism based on two selection
cycles within a scheduling loop. The scheduler will take into account a
scheduling failure and then perform a second scheduling cycle after preemptible
instances have been evacuated ---either by the scheduler itself or by an
external service. However, this two-cycle scheduling mechanism would introduce
a larger scheduling latency and load in the system. This latency is something
perceived negatively by the users \cite{LopezGarcia2016a} so the challenge here
is how to perform this selection in a efficient way, ensuring that the selected
preemptible instances are the less costly for the provider.

\subsection{Preemptible-aware scheduler}
\label{sec:preempt:sched}

\begin{figure}[htbp!]
    \includegraphics[width=\linewidth]{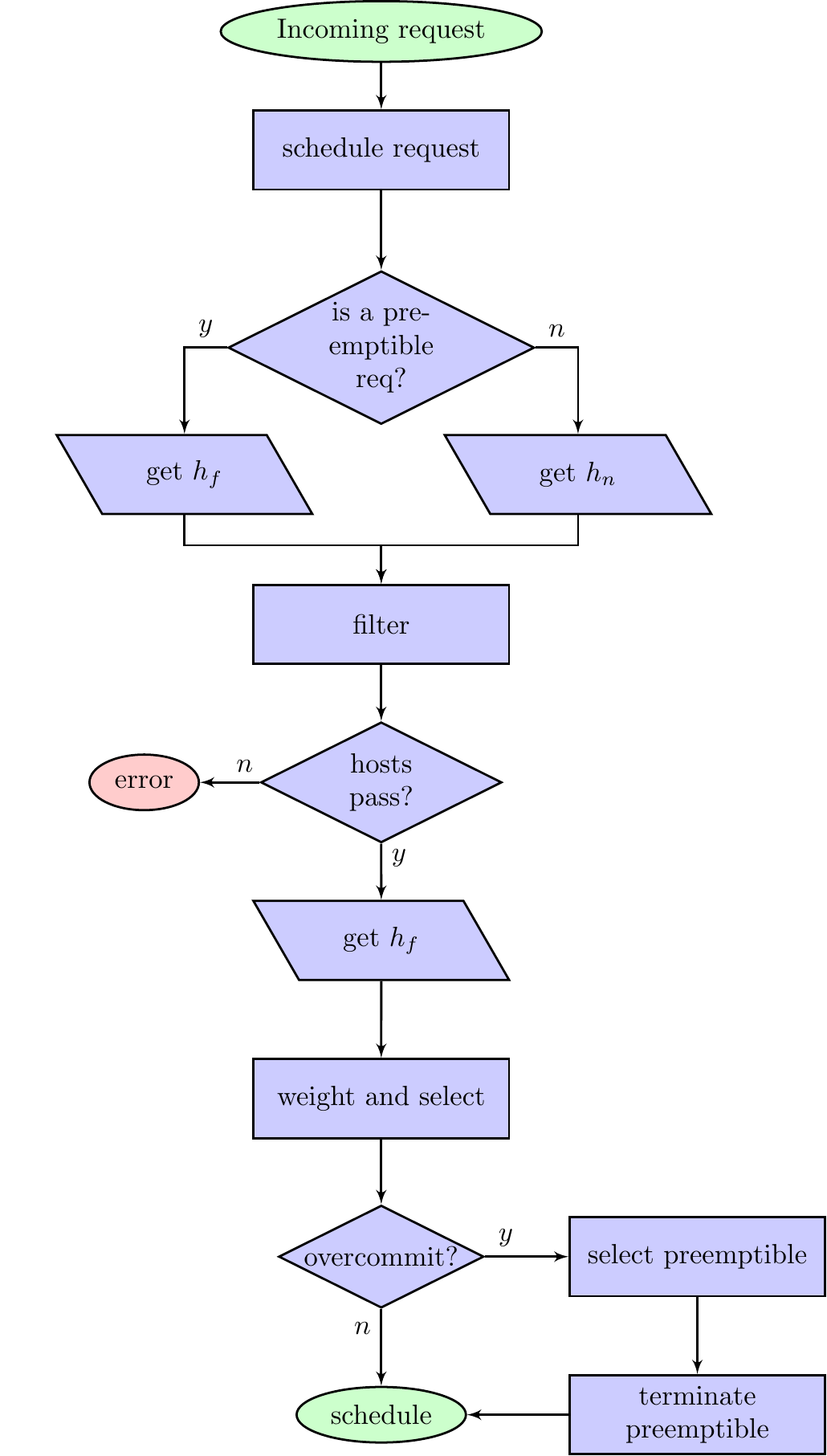}
    \caption{Preemptible Instances Scheduling Algorithm.}
    \label{fig:diagram}
\end{figure}

Our proposed algorithm (depicted in Figure~\ref{fig:diagram}) addresses the
preemptible instances scheduling within one scheduling loop, without
introducing a retry cycle, bur rather performing the scheduling taking into
account different host states depending on the instance that is to be
scheduled. This design takes into account the fact that all the algorithms
described in Section~\ref{sec:sched:algo} are based on two complimentary
phases: filtering and raking., but adds a final phase, where the preemptible
instances that need to be terminated are selected. The algorithm pseudocode is
shown in \ref{alg:schedpreempt} and will be further described in what follows.

As we already explained, the filtering phase eliminates the nodes that are not
able to host the new request due to its current state ---for instance, because
of a lack of resources or a VM anti-affinity---, whereas the raking phase is
the one in charge of assigning a rank or weight to the filtered hosts so that
the best candidate is selected.

\begin{algorithm}
    \caption{Preemptible-aware Scheduling Algorithm.}
    \label{alg:schedpreempt}
    \begin{algorithmic}[1]
        \Function{Select hosts}{$req, H_f, H_n$}
            \INPUT{$req$: user request}
            \INPUT{$H_f$: host full-states}
            \INPUT{$H_n$: host normal-instances states}
            \State {$hosts \gets [~]$} \Comment{empty list}

            \ForAll {$h_{fi}, h_{ni} \in H_f, H_n$}
                \If {\Call{Is Preemptible}{$req$}}
                    \State{$h_i \gets h_{fi}$}
                \Else
                    \State{$h_i \gets h_{ni}$}
                \EndIf

                \If {\Call{Filter}{$h_i, req$}}
                    \State $\Omega_i \gets 0$
                    \ForAll {$r, m$ in ranks} \Comment{$r$ is a
                    rank function, $m$ the rank multiplier}
                        \State $\Omega_i \gets \Omega_i + m_j * r_j(h_{fi}, req) $
                    \EndFor
                    \State $hosts \gets hosts + (h_{fi}, \Omega_i)$ \Comment{append to the list}
                \EndIf
            \EndFor

            \State \textbf{return} $hosts$
        \EndFunction

        \Function{Schedule Request}{$req, H_f, H_n$}
            \INPUT{$req$: user request}
            \INPUT{$H_f$: host full-states}
            \INPUT{$H_n$: host normal-instances states}

            \State {$hosts \gets$} \Call{Select hosts}{$req, H_f, H_n$}
            \State {$host \gets$} \Call{Best Host}{$hosts$}

            \State \Call{Select and Terminate}{$req, host$}

            \State \Return $host$
        \EndFunction
    \end{algorithmic}
\end{algorithm}

I our preemptible-aware scheduler, the filtering phase only takes into account
preemptible instances when doing the filtering phase. In order to do so we
propose to utilize two different states for the physical hosts:

\begin{description}
    \item[$h_f$] This state will take into account all the
        running VM inside that host, that is, the preemptible and non
        preemptible instances.
    \item[$h_n$] This state will not take into
        account all the preemptible instances inside that host. That is, the
        preemptible instances running into a particular physical host are not
        accounted in term of consumed resources.
\end{description}

Whenever a new request arrives, the scheduler will use the $h_f$ or $h_n$ host
states for the filtering phase, depending on the type of the request:

\begin{itemize}
    \item When a normal request arrives, the scheduler will use $h_n$.
    \item When a preemptible request arrives, the scheduler will use $h_f$.
\end{itemize}

This way the scheduler ensures that a normal instance can run regardless of any
preemptible instance occupying its place, as the $h_n$ state does not account
for the resources consumed by any preemptible instance running on the host.
After this stage, the resulting list of hosts will contain all the hosts
susceptible to host the new request, either by evacuating one or several
preemptible instances or because there are enough free resources.

Once the hosts are filtered out, the ranking phase is started. However, in
order to perform the correct ranking, it is needed to use the full state of the
hosts, that is, $h_f$. This is needed as the different rank functions will
require the information about the preemptible instances so as to select the best
node. This list of filtered hosts may contain hosts that are able to accept the
request because they have free resources and nodes that would imply the
termination of one or several instances.

In order to choose the best host for scheduling a new instance new ranking
functions need to be implemented, in order to prioritise the \emph{costless}
host. The simplest ranking function based on the number of preemptible instances
per host is described in Algorithm~\ref{alg:weight:overcommit}.

\begin{algorithm}[h]
    \caption{Ranking function detecting overcommit of resources.}
    \label{alg:weight:overcommit}
    \begin{algorithmic}[1]
        \Function{Overcommit Rank}{$req, h_f$}
            \INPUT{$req$: user request}
            \INPUT{$h_f$: host state}
            \If {$req.resources > h_f.free\_resources$}
                \State \textbf{return} $-1$
            \EndIf
            \State \Return $0$
        \EndFunction
    \end{algorithmic}
\end{algorithm}

This function assigns a negative value if the free resources are not enough to
accommodate the request, detecting an overcommit produced by the fact that it
is needed to terminate one or several preemptible instances. However, this
basic function only establishes a naive ranking based on the termination or not
of instances. In the case that it is needed to terminate various instances,
this function does not establish any rank between them, so more appropriate
rank functions need to be created, depending on the business model implemented
by the provider. Our design takes this fact into account, allowing for
modularity of these cost functions that can be applied to the raking function.

For instance, commercial providers tend to charge by complete periods of
\SI{1}{\hour}, so partial hours are not accounted. A ranking function based in
this business model can be expressed as Algorithm~\ref{alg:weight:mod}, ranking
hosts according to the preemptible instances running inside them and the time
needed until the next complete period.

\begin{algorithm}[h]
    \caption{Ranking function based on \SI{1}{\hour} consumption periods.}
    \label{alg:weight:mod}
    \begin{algorithmic}[1]
        \Function{Period Rank}{$req, h_f$}
            \INPUT{$req$: user request}
            \INPUT{$h_f$: host state}
            \State {$weight \gets 0$}
            \ForAll {$instance \in get\_instances(h_f)$}
                \If {$(is\_spot(instance)$}
                    \If {$(instance.run\_time \mod 3600) > 0$}
                        \State {$weight \gets weight + instance.run\_time \mod 3600$}
                    \EndIf
                \EndIf
            \EndFor
            \State \Return $- weight$
        \EndFunction
    \end{algorithmic}
\end{algorithm}

Once the ranking phase is finished, the scheduler will have built an ordered
list of hosts, containing the best candidates for the new request. Once the
best host selected it is still needed to select which individual preemptible
instances need to be evacuated from that host, if any. Our design adds a
third phase, so as to terminate the preemptible instances if needed.

This last phase will perform an additional raking and selection of the
candidate preemptible instances inside the selected host, so as to select the
less costly for the provider. This selection leverages a similar ranking
process, performed on the preemptible instances, considering all the
preemptible instances combination and the costs for the provider, as shown in
Algorithm~\ref{alg:selection}.

\begin{algorithm}[h]
    \caption{Preemptible instance selection and termination.}
    \label{alg:selection}
    \begin{algorithmic}[1]
        \Procedure{Select and Terminate}{$req, h_f$}
            \INPUT{$req$: user request}
            \INPUT{$h_f$: host state}
            \State {$selected\_instances \gets [~]$}
            \ForAll {$instances \in get\_\_all\_preemptible\_combinations(h_f)$}
                \If {$\sum(instances.resources) > req.resources$}
                    \If {$cost(instances) < cost(selected\_instances)0$}
                        \State {$selected\_instances \gets instances$}
                    \EndIf
                \EndIf
            \EndFor
            \State \Call{Terminate}{selected\_instances}
        \EndProcedure
    \end{algorithmic}
\end{algorithm}

\section{Evaluation}
\label{sec:evaluation}

In the first part of this section (\ref{sec:evaluation:implementation}) we will
describe an implementation ---done for the OpenStack Compute CMF---, in order
to evaluate our proposed algorithm. We have decided to implement it on top of
the OpenStack Compute software due to its modular design, that allowed us to
easily plug our modified modules without requiring significant modifications to
the code core.

Afterwards we will perform two different evaluations. On the one hand we will
assess the algorithm correctness, ensuring that the most desirable instances
are selected according to the configured weighers
(Section~\ref{sec:evaluation:algorithm}). On the other hand we will examine the
performance of the proposed algorithm when compared with the default scheduling
mechanism (Section~\ref{sec:evaluation:usecase}).

\subsection{OpenStack Compute Filter Scheduler}

The OpenStack Compute scheduler is called Filter Scheduler and, as already
described in Section~\ref{sec:related}, it is a rank scheduler, implementing
two different phases: filtering and weighting.

\begin{description}
    \item[Filtering] The first step is the filtering phase. The scheduler
        applies a concatenation of filter functions to the initial set of
        available hosts, based on the host properties and state ---e.g. free
        RAM or free CPU number--- user input ---e.g. affinity or anti-affinity
        with other instances--- and resource provider defined configuration.
        When the filtering process has concluded, all the hosts in the final
        set are able to satisfy the user request.

    \item[Weighing] Once the filtering phase returns a list of suitable hosts,
        the weighting stage starts so that the best host ---according to the
        defined configuration--- is selected.  The scheduler will apply all
        hosts the same set of weigher functions $\mathrm{w}_i(h)$, taking
        into account each host state $h$. Those weigher functions will return
        a value considering the characteristics of the host received as input
        parameter, therefore, total weight $\Omega$ for a node $h$ is
        calculated as follows:

        $$
        \Omega = \sum^n{m_i\cdot \mathrm{N}{(\mathrm{w}_i(h))}}
        $$

        Where $m_i$ is the multiplier for a weigher function,
        $\mathrm{N}{(\mathrm{w}_i(h))}$ is the
        normalized weight between $[0, 1]$ calculated via a rescaling like:

        $$
        \mathrm{N}{(\mathrm{w}_i(h))} = \frac{\mathrm{w}_i(h)-\min{W}}{\max{W} - \min{W}}
        $$

        where $\mathrm{w}_i(h)$ is the weight function, and $\min{W}$,
        $\max{W}$ are the minimum and maximum values that the weigher has
        assigned for the set of weighted hosts. This way, the final weight
        before applying the multiplication factor will be always in the range
        $[0, 1]$.
\end{description}

After these two phases have ended, the scheduler has a set of hosts, ordered
according to the weights assigned to them, thus it will assign the request to
the host with the maximum weight. If several nodes have the same weight, the
final host will be randomly selected from that set.

\subsection{Implementation Evaluation}
\label{sec:evaluation:implementation}

We have extended the Filter Scheduler algorithm with the functionality described
in Algorithm~\ref{alg:spot}. We have also implemented the ranking functions
described in Algorithm~\ref{alg:weight:overcommit} and
Algorithm~\ref{alg:weight:mod} as weighers, using the OpenStack terminology.

\begin{algorithm}[h!]
    \caption{Preemptible Instances Scheduling Algorithm.}
    \label{alg:spot}
    \begin{algorithmic}[1]
        \Function{Select Destinations}{$req$}
        \INPUT{$req$: user request}
            \State {$host \gets \Call{Schedule}{req}$}
            \If {$host$ is overcommitted}
                \State \Call{Select and Terminate}{req,host}
            \EndIf
            \State \textbf{return} $host$
        \EndFunction
        \Statex
        \Function{Schedule}{$req$}
            \INPUT{$req$: user request}
            \State {$H_f \gets host\_states(full)$}
            \State {$H_p \gets host\_states(partial)$}
            \If{$is\_spot(req)$}
                \State {$H_{filtered} \gets filter(req, H_f)$}
            \Else
                \State {$H_{filtered} \gets filter(req, H_p)$}
            \EndIf
            \State {$H_{weighted} \gets weight(req, H_f)$}
            \State $best \gets select\_best(H_{weighted})$
            \State \textbf{return} $best$
        \EndFunction
        \Statex
        \Procedure{Select and Terminate}{$req, h_f$}
            \INPUT{$req$: user request}
            \INPUT{$h_f$: host state}
            \State {$selected\_instances \gets [~]$}
            \ForAll {$instances \in get\_all\_preemptible\_combinations(h_f)$}
                \If {$\sum{instances.resources} > req.resources$}
                    \If {$cost(instances) < cost(selected\_instances)0$}
                        \State {$selected\_instances \gets instances$}
                    \EndIf
                \EndIf
            \EndFor
            \State \Call{Terminate}{selected\_instances}
        \EndProcedure
    \end{algorithmic}
\end{algorithm}

Moreover, the Filter Scheduler has been also modified so as to introduce the
additional select and termination phase (Algorithm~\ref{alg:selection}). This
phase has been implemented following the same same modular approach as the
OpenStack weighting modules, allowing to define and implement additional cost
modules to determine which instances are to be selected for termination.

As for the cost functions, we have implemented a module following
Algorithm~\ref{alg:weight:mod}. This cost function assumes that customers are
charged by periods of \SI{1}{\hour}, therefore it prioritizes the termination
of Spot Instances with the lower partial-hour consumption (i.e. if we consider
instances with \SI{120}{\minute}, \SI{119}{\minute} and \SI{61}{\minute} of
duration, the instance with \SI{120}{\minute} will be terminated).

This development has been done on the OpenStack Newton
version\footnote{\url{https://github.com/indigo-dc/opie}}, and was deployed on
the infrastructure that we describe in Section~\ref{sec:evaluation:config}.

\subsection{Configurations}
\label{sec:evaluation:config}

In order to evaluate our algorithm proposal we have set up a dedicated test
infrastructure comprising a set of \num{26} identical IBM HS21 blade servers,
with the characteristics described in Table~\ref{tb:ifcatest}. All the nodes
had an identical base installation, based on an Ubuntu Server 16.04 LTS,
running the Linux 3.8.0 Kernel, where we have deployed OpenStack Compute as the
Cloud Management Framework. The system architecture is as follows:

\begin{itemize}
    \item A Head node hosting all the required services to manage the
        cloud test infrastructure, that is:
        \begin{itemize}
            \item The OpenStack Compute API.
            \item The OpenStack Compute Scheduler service.
            \item The OpenStack Compute Conductor service.
            \item The OpenStack Identity Service (Keystone)
            \item A MariaDB 10.1.0 server.
            \item A RabbitMQ 3.5.7 server.
        \end{itemize}
    \item An Image Catalog running the OpenStack Image Service (Glance) serving
        images from its local disk.
    \item \num{24} Compute Nodes running OpenStack Compute, hosting the spawned
        instances.
\end{itemize}

\begin{table}[!t]
	\renewcommand{\arraystretch}{1.3}
    \caption{Test node characteristics.}
    \label{tb:ifcatest}
	\centering
    \begin{tabular}{r|l}

        \textbf{CPU}     & \num{2} x Intel\textregistered Xeon\textregistered Quad Core E5345 \SI{2.33}{\giga\hertz}\\
		\hline
        \textbf{RAM}     & \SI{16}{\giga\byte}\\
		\hline
        \textbf{Disk}    & \SI{140}{\giga\byte}, \num{10000} rpm hard disk \\
		\hline
        \textbf{Network} & \SI{1}{\giga\bit} Ethernet\\
    \end{tabular}
\end{table}

The network setup of the testbed consists on two \SI{10}{\giga\bit} Ethernet
switches, interconnected with a \SI{10}{\giga\bit} Ethernet link. All the hosts
are evenly connected to these switches using a \SI{1}{\giga\bit} Ethernet
connection.

We have considered the VM sizes described in Table~\ref{tab:sizes}, based on
the default set of sizes existing in a default OpenStack installation.

\begin{table}[!t]
	\renewcommand{\arraystretch}{1.3}
    \caption{Configured VM sizes.}
    \label{tab:sizes}
	\centering
    \begin{tabular}{|r|c|c|c|}
		\hline
        \bfseries Name & \bfseries vCPUs & \bfseries RAM (\si{\mega\byte}) & \bfseries Disk (\si{\giga\byte})\\
		\hline
		\hline
        small  & 1     & 2000 &  20   \\
		\hline
        medium & 2     & 4000 &  40   \\
		\hline
        large  & 4     & 8000 &  80   \\
		\hline
    \end{tabular}
\end{table}

\subsection{Algorithm Evaluation}
\label{sec:evaluation:algorithm}

The purpose of this evaluation is to ensure that the proposed algorithm is
working as expected, so that:

\begin{itemize}
    \item The scheduler is able to deliver the resources for a normal request,
        by terminating one or several preemptible instances when there are not
        enough free idle resources.
    \item The scheduler selects the best preemptible instance for termination,
        according to the configured policies by means of the scheduler
        weighers.
\end{itemize}

\subsubsection{Scheduling using same Virtual Machine sizes}

For the first batch of tests, we have considered same size instances, to
evaluate if the proposed algorithm chooses the best physical host and
selects the best preemptible instance for termination. We generated requests for
both preemptible and normal instances ---chosen randomly---, of random duration
between \SI{10}{\minute} and \SI{300}{\minute}, using an exponential
distribution \cite{knuth1981art} until the first scheduling failure for a normal
instance was detected.

The compute nodes used have \SI{16}{\giga\byte} of RAM and eight CPUs, as
already described. The VM size requested was the \emph{medium} one, according
to Table~\ref{tab:sizes}, therefore each compute node could host up to four
VMs.

We executed these requests and monitored the infrastructure until the first
scheduling failure for a normal instance took place, thus the preemptible
instance termination mechanism was triggered. At that moment we took a snapshot
of the nodes statuses, as shown in Table~\ref{tab:evaluation1} and
Table~\ref{tab:evaluation2}. These tables depict the status for each of
the physical hosts, as well as the running time for each of the instances that
were running at that point. The shaded cells represents the preemptible
instance that was terminated to free the resources for the incoming non
preemptible request.

\begin{table}[!t]
	\renewcommand{\arraystretch}{1.3}
    \caption{Test-1, preemptible instances evaluation using the same VM
        size. The label marked with (\protect\footnotemark[1]) indicate the
        terminated instance. Time is expressed in minutes.}
    \label{tab:evaluation1}
    \begin{minipage}[c]{\linewidth}
    \renewcommand\footnoterule{}
    \renewcommand{\thefootnote}{\alph{footnote}}
	\centering
    \begin{tabular}{|c|c|c|c|c|}
		\hline
        \bfseries \multirow{2}{*}{Host}   & \multicolumn{2}{c|}{\bfseries Instances} & \multicolumn{2}{c|}{\bfseries Preeptible Instances} \\
        \cline{2-5}
                                & \bfseries ID & \bfseries Time & \bfseries ID & \bfseries Time \\
		\hline
		\hline
        \multirow{2}{*}{host-A} & A1 & 272 & AP1 & 96 \\
                                & A2 & 172 & AP2 & 207 \\
		\hline
        \multirow{2}{*}{host-B} & B1 & 136 & BP1 (\footnotemark[1]) & 71 \\
                                & B2 & 200 & BP2 & 91 \\
		\hline
        \multirow{2}{*}{host-C} & C1 &  97 & CP1 & 210 \\
                                & C2 & 275 & CP2 & 215 \\
		\hline
        \multirow{3}{*}{host-D} & \multirow{3}{*}{D1} & \multirow{3}{*}{16} & DP1 & 85 \\
                                &    &    & DP2 & 199 \\
                                &    &    & DP3 & 152 \\
		\hline
    \end{tabular}
    \footnotetext[1]{Selected instance}
    \end{minipage}
\end{table}

Considering that the preemptible instance selection was done according to
Algorithm~\ref{alg:selection} using the cost function in
Algorithm~\ref{alg:weight:mod}, the chosen instance has to be the one with the
lowest partial-hour period. In Table~\ref{tab:evaluation1} this is the instance
marked with (\footnotemark[1]): \emph{BP1}. By chance, it corresponds with the
preemptible instance with the lowest run time.

\begin{table}[!t]
	\renewcommand{\arraystretch}{1.3}
    \caption{Test-2, preemptible instances evaluation using the same VM
            size. The label marked with (\protect\footnotemark[1]) indicate the
            terminated instance. Time is expressed in minutes.}
    \label{tab:evaluation2}
    \begin{minipage}[c]{\linewidth}
    \renewcommand\footnoterule{}
    \renewcommand{\thefootnote}{\alph{footnote}}
	\centering
    \begin{tabular}{|c|c|c|c|c|}
		\hline
        \bfseries \multirow{2}{*}{Host}   & \multicolumn{2}{c|}{\bfseries Instances} & \multicolumn{2}{c|}{\bfseries Preeptible Instances} \\
        \cline{2-5}
                                & \bfseries ID & \bfseries Time & \bfseries ID & \bfseries Time \\
		\hline
		\hline
        \multirow{4}{*}{host-A} &    &     & AP1 & 247 \\
                                &    &     & AP2 & 463 \\
                                &    &     & AP3 & 403 \\
                                &    &     & AP4 & 410 \\
		\hline
        \multirow{2}{*}{host-B} & B1 & 388 & BP1 & 344 \\
                                & B2 & 103 & BP2 & 476 \\
		\hline
        \multirow{2}{*}{host-C} & C1 & 481 & CP1 (\footnotemark[1]) & 181 \\
                                & C2 & 177 & CP2 & 160 \\
		\hline
        \multirow{3}{*}{host-D} & \multirow{3}{*}{D1} & \multirow{3}{*}{173} & DP1 & 384 \\
                                &    &    & DP2 & 168 \\
                                &    &    & DP3 & 232 \\
		\hline
    \end{tabular}
    \footnotetext[1]{Selected instance.}
    \end{minipage}
\end{table}

Table~\ref{tab:evaluation2} shows a different test execution under the
same conditions and constraints. Again, the selected instance has to be the one
with the lowest partial-hour period. In Table~\ref{tab:evaluation2} this
corresponds to the instance marked again with (\footnotemark[1]): \emph{CP1}, as
its remainder is \SI{1}{\minute}. In this case this is not the preemptible
instance with the lowest run time (being it \emph{CP2}).

\subsubsection{Scheduling using different Virtual Machine sizes}

For the second batch of tests we requested instances using different sizes,
always following the sizes in Table~\ref{tab:sizes}.
Table~\ref{tab:evaluation3} depicts the testbed status when a request for a
\emph{large} VM caused the termination of the instances marked with
(\footnotemark[1]): \emph{AP2}, \emph{AP3} and \emph{AP4}. In this case, the
scheduler decided that the termination of these three instances caused a
smaller impact on the provider, as the sum of their \SI{1}{\hour} remainders
(\num{55}) was lower than any of the other possibilities (\num{58} for
\emph{BP1}, \num{57} for \emph{CP1}, \num{112} for \emph{CP2} and \emph{CP3}).

\begin{table}[!t]
	\renewcommand{\arraystretch}{1.3}
    \caption{Test-3, preemptible instances evaluation using different VM sizes.
        The labels marked with (\protect\footnotemark[1]) indicate the
        terminated instances. Time is
        expressed in minutes. S, M, L stand for Small, Medium and Large
        respectively.}
    \label{tab:evaluation3}
    \begin{minipage}[c]{\linewidth}
    \renewcommand\footnoterule{}
    \renewcommand{\thefootnote}{\alph{footnote}}
    \centering
    \begin{tabular}{|c|c|c|c|c|c|c|}
        \hline
        \bfseries \multirow{2}{*}{Host}   & \multicolumn{3}{c|}{\bfseries Instances} & \multicolumn{3}{c|}{\bfseries Preeptible Instances} \\
        \cline{2-7}
                                & \bfseries ID & \bfseries Time & \bfseries Size & \bfseries ID & \bfseries Time& \bfseries Size \\
        \hline
        \hline
        \multirow{2}{*}{host-A} &    &    & & AP1 & 298 & L \\
                                &    &    & & AP2 (\protect\footnotemark[1]) & 278 & M \\
                                &    &    & & AP3 (\protect\footnotemark[1]) & 190 & S \\
                                &    &    & & AP4 (\protect\footnotemark[1]) & 187 & S \\
        \hline
        \multirow{1}{*}{host-B} & B1 & 494 & L & BP1 & 178 & L \\
        \hline
        \multirow{3}{*}{host-C} &    &     & & CP1 & 297 & L \\
                                &    &     & & CP2 & 296 & M \\
                                &    &     & & CP3 & 296 & S \\
        \hline
        \multirow{3}{*}{host-D} & D1 & 176 & M &     &    & \\
                                & D2 & 200 & M &     &    & \\
                                & D3 & 116 & L &     &    & \\
        \hline
    \end{tabular}
    \footnotetext[1]{Selected instances.}
    \end{minipage}
\end{table}

Table~\ref{tab:evaluation4} shows a different test execution under the same
conditions and constraints. In this case, the preemptible instance termination
was triggered by a new VM request of size \emph{medium} and the selected
instance was the one marked with (\protect\footnotemark[1]): \emph{BP3}, as
\emph{host-B} will have enough free space just by terminating one instance.

\begin{table}[!t]
	\renewcommand{\arraystretch}{1.3}
    \caption[Test-4, preemptible instances evaluation using different VM sizes.]
            {Test-4, preemptible instances evaluation using different VM sizes.
                The labels marked with (\protect\footnotemark[1]) indicate the
                terminated instances. Time is
                expressed in minutes. S, M, L stand for Small, Medium and Large
                respectively.}
    \label{tab:evaluation4}
    \begin{minipage}[c]{\linewidth}
    \renewcommand\footnoterule{}
    \renewcommand{\thefootnote}{\alph{footnote}}
    \centering
    \begin{tabular}{|c|c|c|c|c|c|c|}
        \hline
        \bfseries \multirow{2}{*}{Host}   & \multicolumn{3}{c|}{\bfseries Instances} & \multicolumn{3}{c|}{\bfseries Preeptible Instances} \\
        \cline{2-7}
                                & \bfseries ID & \bfseries Time & \bfseries Size & \bfseries ID & \bfseries Time& \bfseries Size \\
        \hline
        \hline
        \multirow{2}{*}{host-A} & A1 & 234 & L & AP1 & 172 & M \\
                                & A2 & 122 & M &     &    & \\
        \hline
        \multirow{3}{*}{host-B} &    &    & & BP1 & 272 & L \\
                                &    &    & & BP2 & 212 & M \\
                                &    &    & & BP3 (\protect\footnotemark[1]) & 380 & S \\
        \hline
        \multirow{3}{*}{host-C} & C1 & 182 & S &     &    & \\
                                & C2 & 120 & M &     &    & \\
                                & C3 & 116 & L &     &    & \\
        \hline
        \multirow{4}{*}{host-D} &    &     & & DP1 & 232 & L \\
                                &    &     & & DP2 & 213 & S \\
                                &    &     & & DP3 & 324 & M \\
                                &    &     & & DP4 & 314 & S \\
        \hline
    \end{tabular}
    \footnotetext[1]{Selected instances.}
\end{minipage}
\end{table}

\subsection{Performance evaluation}
\label{sec:evaluation:usecase}

As we have already said in Section~\ref{sec:design}, we have focused on
designing an algorithm that does not introduce a significant latency in the
system. This latency will introduce a larger delay when delivering the
requested resources to the end users, something that is not desirable by any
resource provider \cite{LopezGarcia2017c}.

In order to evaluate the performance of our proposed algorithm we have done a
comparison with the default, unmodified OpenStack Filter Scheduler. Moreover,
for the sake of comparison, we have implemented a scheduler based on a retry
loop as well. This scheduler performs a normal scheduling loop, and if there is
a scheduling failure for a normal instance, it will perform a second pass
taking into account the existing preemptible instances. The preemptible
instance selection and termination mechanisms remain the same.

We have scheduled 130 Virtual Machines of the same size on our test
infrastructure and we have recorded the timings for the scheduling function,
thus calculating the means and standard deviation for each of the following
scenarios:

\begin{itemize}
    \item Using the original, unmodified OpenStack Filter scheduler with an
        empty infrastructure.
    \item Using the preemptible instances Filter Scheduler and the retry
        scheduler:
        \begin{itemize}
            \item Requesting normal instances with an empty infrastructure.
            \item Requesting preemptible instances with an empty infrastructure.
            \item Requesting normal instances with a saturated infrastructure,
                thus implying the termination of a preemptible instance each
                time a request is performed.
        \end{itemize}
\end{itemize}

\begin{figure}[!t]
	\centering
    \includegraphics[height=0.95\linewidth]{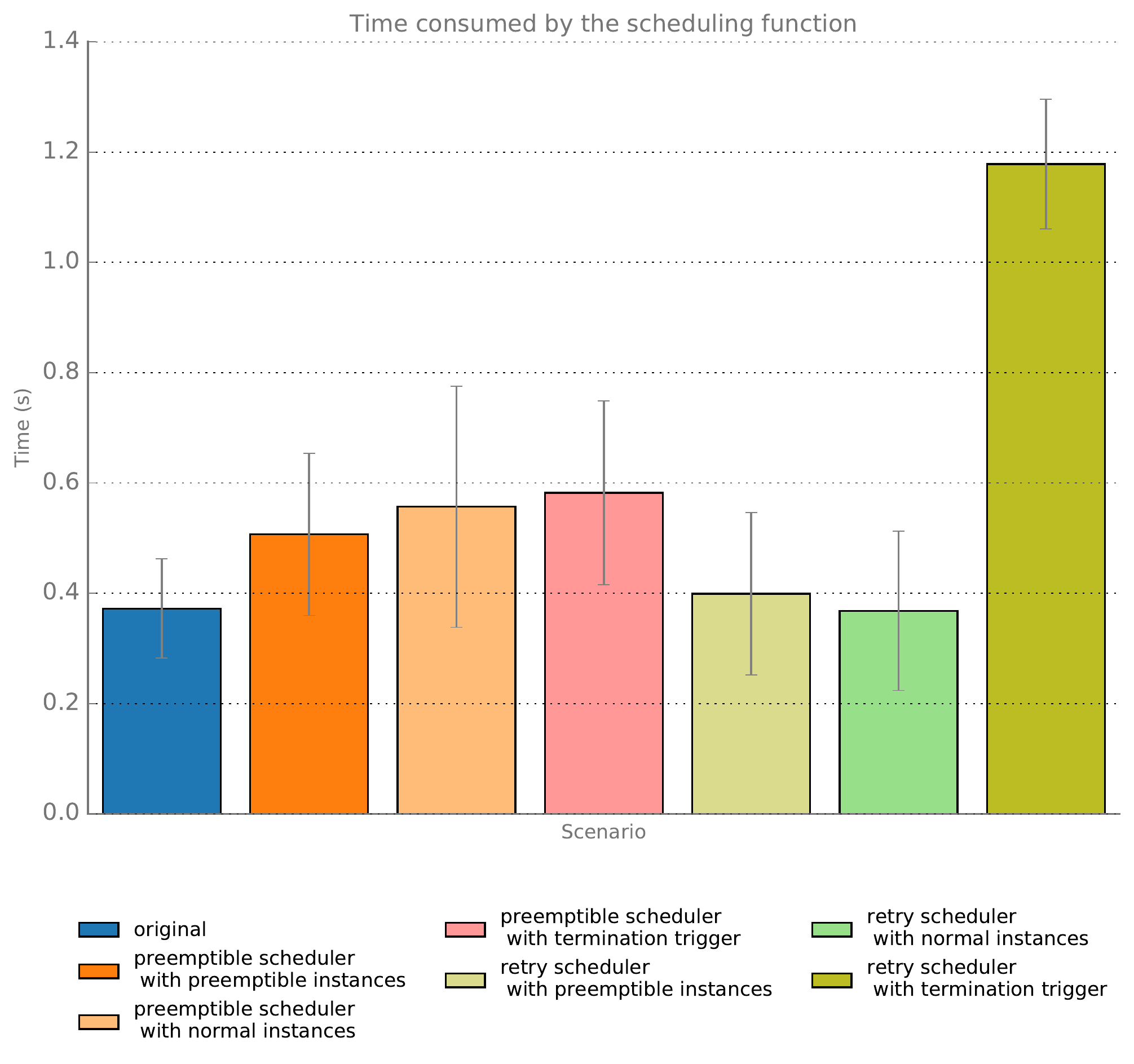}
	\caption{Comparison of the time consumed by the different scheduling
options in different scenarios. Error bars represent the standard deviation.}
    \label{fig:comparison}
\end{figure}

We have then collected the scheduling calls timings and we have calculated the
means and deviations for each scenario, as shown in
Figure~\ref{fig:comparison}. Numbers in these scenarios are quite low, since
the infrastructure is a small testbed, but these numbers are expected to become
larger as the infrastructure grows in size.

As it can be seen in the
aforementioned Figure~\ref{fig:comparison}, our solution introduces a delay
in the scheduling calls, as we need to calculate additional host
states (we hold two different states for each node) and we need to select a
preemptible instance for termination (in case it is needed). In the case of the
retry scheduler, this delay does not exists and numbers are similar to the
original scheduler. However, when it is needed to trigger the termination of a
preemptible instance, having a retry mechanism (thus executing the same
scheduling call two times) introduces a significantly larger penalty when
compared to our proposed solution.
We consider that the latency that we are introducing is within an acceptable
range, therefore not impacting significantly the scheduler performance.

\section{Exploitation and integration in existing infrastructures}

The functionality introduced by the preemptible instances model that we have
described in this work can be exploited not only within a cloud resource
provider, but it can also be leveraged on more complex hybrid infrastructures.

\subsection{High Performance Computing Integration}
\label{sec:additional}

One can find in the literature several exercises of integration of hybrid
infrastructures, integrating cloud resources, commercial or private, with High
Performance Computing (HPC) resources. Those efforts focus on outbursting
resources from the cloud, when the HPC system does not provide enough resources
to solve a particular problem \cite{EC2-HPC}.

On-demand provisioning using cloud resources when the batch system of the HPC
is full is certainly a viable option to expand the capabilities of a HPC center
for serial batch processing.

We focus however in the complementary approach, this is, using HPC resources to
provide cloud resources capability, so as to complement existing distributed
infrastructures. Obviously HPC systems are oriented to batch processing of
highly coupled (parallel) jobs. The question here is optimizing resource
utilization when the HPC batch system has empty slots.

If we backfill the empty slots of a HPC system with cloud jobs, and a new
regular batch job arrives from the HPC users, the cloud jobs occupying the
slots needed by the newly arrived batch job should be terminated immediately,
so as to not disturb regular work. Therefore such cloud jobs should be
submitted as Spot Instances

Enabling HPC systems to process other jobs during periods in which the load of
the HPC mainframe is low, appears as an attractive possibility from the point
of view of resource optimization. However the practical implementation of such
idea would need to be compatible with both, the HPC usage model, and the cloud
usage model.

In HPC systems users login via ssh to a frontend. At the frontend the user has
the tools to submit jobs. The scheduling of HPC jobs is done using a regular
batch systems software (such as SLURM, SGE, etc...).

HPC systems are typically running MPI parallel jobs as well using specialized
hardware interconnects such as Infiniband.

Let us imagine a situation in which the load of the HPC system is low. One can
instruct the scheduler of the batch system to allow cloud jobs to  HPC system
occupying those slots not allocated by the regular batch allocation.

In order to be as less disrupting as possible the best option is that the cloud
jobs arrive as preemptible instances as described through this paper.  When a
batch job arrives to the HPC system, this job should be immediately scheduled
and executed. Therefore the scheduler should be able to perform the following
steps:

\begin{itemize}
    \item{} Allocate resources for the job that just arrived to the batch queue
        system
    \item{} Identify the cloud jobs that are occupying those resources, and
        stop them.
    \item{} Dispatch the batch job.
\end{itemize}

In the case of parallel jobs the scheduling decision may depend on many factors
like the topology of the network requested, or the affinity of the processes at
the core/CPU level. In any case parallel jobs using heavily the low latency
interconnect should not share nodes with any other job.

\subsection{High Throughput Computing Integration}

Existing High Throughput Computing Infrastructures, like the service offered by
EGI\footnote{\url{https://www.egi.eu/services/high-throughput-compute/}}, could benefit from a cloud providers offering preemptible
instances. It has been shown that cloud resources and IaaS offerings can be
used to run HTC tasks \cite{McNab2014} in a pull mode, where cloud instances
are started in a way that they are able to pull computing tasks from a central
location (for example using a distributed batch system like HTCondor).

However, sites are reluctant to offer large amounts of resources to be used in
this mode due to the lack of a fixed duration for cloud instances.
In this context, federated cloud e-Infrastrucutres like the EGI Federated Cloud
\cite{FernandezdelCastillo2015}, could benefit from resource providers offering
preemptible instances. Users could populate idle resources with preemptible
instances pulling their HTC tasks, whereas interactive and normal IaaS users
will not be impacted negatively, as they will get the requests satisfied. In
this way, large amounts of cloud computing power could be offered to the
European research community.

\section{Conclusions}
\label{sec:conclusions}

In this work we have proposed a preemptible instance scheduling design that
does not modify substantially the existing scheduling algorithms, but rather
enhances them. The modular rank and cost mechanisms allows the definition and
implementation of any resource provider defined policy by means of additional
pluggable rankers. Our proposal and implementation enables all kind of service
providers ---whose infrastructure is managed by open source middleware such as
OpenStack--- to offer a new access model based on preemptible instances, with a
functionality similar to the one offered by the major commercial providers.

We have checked for the algorithm correctness when selecting the preemptible
instances for termination. The results yield that the algorithm behaves as
expected. Moreover we have compared the scheduling performance with regards
equivalent default scheduler, obtaining similar results, thus ensuring that the
scheduler performance is not significantly impacted.

This implementation allows to apply more complex policies on top of the
preemptible instances, like instance termination based on price fluctuations
(that is, implementing a preemptible instance stock market), preemptible
instance migration so as to consolidate them or proactive instance termination
to maximize the provider's revenues by not delivering computing power at no
cost to the users.

\section*{Acknowledgements}

The authors acknowledge the financial support from the European Commission
Horizon 2020 via INDIGO-DataCloud project (grant number 653549) and EGI-ENGAGE
(grant number 654142) and the Ministry of Economy and Competitiveness for the
support through the National Plan under contract number FPA2013-40715-P.

The authors want also to thank the IFCA Advanced Computing and e-Science Group.

\bibliographystyle{elsarticle-num}
\bibliography{references}

\end{document}